\documentclass[twocolumn,showpacs,amsmath,amssymb,superscriptaddress]{revtex4}
\usepackage{dcolumn}
\usepackage{color}
\usepackage[dvips]{graphicx}
\usepackage{amsmath}
\usepackage{bm}

\begin{document}

\title{Microscopic analysis of the octupole phase transition in Th isotopes}

\author{K.~Nomura}
\affiliation{Institut f\"ur Kernphysik, Universit\"at zu K\"oln, D-50937
K\"oln, Germany}
\affiliation{Physics Department, Faculty of Science, University of
Zagreb, 10000 Zagreb, Croatia}

\author{D.~Vretenar}
\affiliation{Physics Department, Faculty of Science, University of
Zagreb, 10000 Zagreb, Croatia}

\author{B.-N. Lu}
\affiliation{State Key Laboratory of Theoretical Physics, Institute of Theoretical Physics, Chinese Academy of Sciences, Beijing 100190, China}
\affiliation{Physics Department, Faculty of Science, University of
Zagreb, 10000 Zagreb, Croatia}

\date{\today}

\begin{abstract}
A shape phase transition between stable octupole deformation and octupole 
vibrations in Th nuclei is analyzed in a microscopic framework based 
on nuclear density functional theory. 
The relativistic functional DD-PC1 is 
used to calculate axially-symmetric quadrupole-octupole constrained energy surfaces. 
Observables related to order parameters are computed using an 
interacting-boson Hamiltonian, with 
parameters determined by mapping the microscopic energy surfaces 
to the expectation value of the Hamiltonian in the boson condensate. 
The systematics of constrained energy surfaces and 
low-energy excitation spectra point to the occurrence of a phase 
transition between octupole-deformed shapes and shapes
characterized by octupole-soft potentials.
\end{abstract}



\maketitle

The evolution of equilibrium shapes and the corresponding excitation dynamics 
present one of the most intriguing aspects of the nuclear many-body system \cite{BM,RS,CasBook}. 
The simplest low-energy collective excitations correspond to quadrupole modes,
that is, the geometrical shape of a nucleus varies between a sphere and a 
rotational ellipsoid. Most deformed nuclei display quadrupole reflection-symmetric 
equilibrium shapes, and the corresponding excitation spectra are characterized by 
positive-parity rotational bands. There are, however, regions of the mass table 
in which octupole deformations (reflection-asymmetric, pear-like shapes) 
occur \cite{butler96}. Reflection-asymmetric shapes are distinguished by the presence 
of negative-parity bands, and by pronounced electric dipole and octupole
transitions. Structure phenomena related to reflection-asymmetric nuclear shapes 
have been explored in numerous studies \cite{scholten78,naza84b,bonche86,engel87,taka88,kusnezov88,butler96,cottle98,bizzeti04,bonatsos05,lenis06,bizzeti08,bizzeti10,jolos12,rayner12,minkov12}.
Analogous excitation patterns are also observed in other mesoscopic
systems like molecules and, therefore, studies of octupole collective
degrees of freedom are of broad interest in many aspects of finite quantum systems. 

The transition between different nuclear shapes in most isotopic or isotonic sequences is
gradual. In some cases, however, the addition/subtraction of only a few nucleons leads to 
rather rapid changes in equilibrium shapes and, in particular, shape 
phase transitions and critical-point phenomena may occur. 
Phase transitions in the equilibrium shapes of nuclei correspond 
to first- and second-order quantum phase transitions (QPTs) between 
competing ground-state phases induced by variation of a non-thermal 
control parameter (number of nucleons) at zero 
temperature \cite{cejnar10rev}. Important issues in studies of 
nuclear QPT include the identification of observables that can be 
related to order parameters, the degree to which discontinuities at a phase transitional
point are smoothed out in finite nuclei, and the question of how precisely can a point of phase
transition be associated with a particular isotope, considering that the control parameter,
i.e. nucleon number, is not continuous but takes only discrete integer values. 
In the last decade nuclear QPTs have been investigated extensively, 
both in experimental studies and employing a variety of theoretical models 
(cf. Ref. \cite{cejnar10rev} for a recent review). 
Most studies have been focused on quadrupole shape phase transitions, but several 
phenomenological models have also considered possible phase transitions 
related to octupole shapes 
\cite{scholten78,kusnezov88,bizzeti04,bonatsos05,lenis06,bizzeti08,bizzeti10,jolos12}.

In this work we analyze shape QPTs in octupole deformed nuclei and
present the first microscopic realization of a QPT from 
stable octupole deformation to octupole vibrations in the Th isotopic chain, characteristic 
for the region of light actinides. 
This study is based on the microscopic framework of nuclear energy density functionals, and a  
corresponding interacting boson model \cite{IBM} Hamiltonian  
is constructed to calculate the excitation spectra and observables that can be
related to quantum order parameters. 
 
\begin{figure*}[ctb!]
\begin{center}
\begin{tabular}{cccc}
\includegraphics[width=5.7cm]{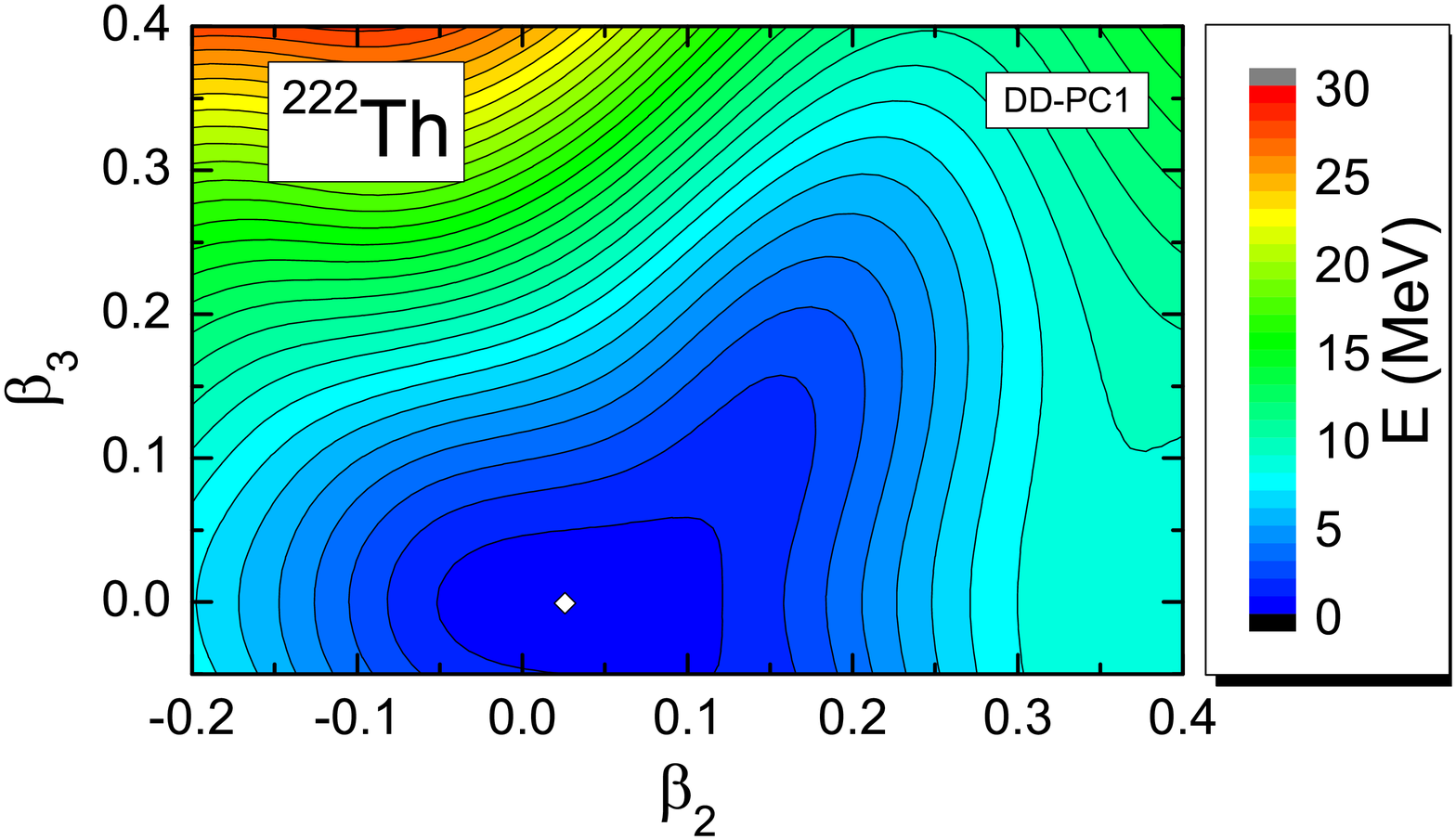} &
\includegraphics[width=5.7cm]{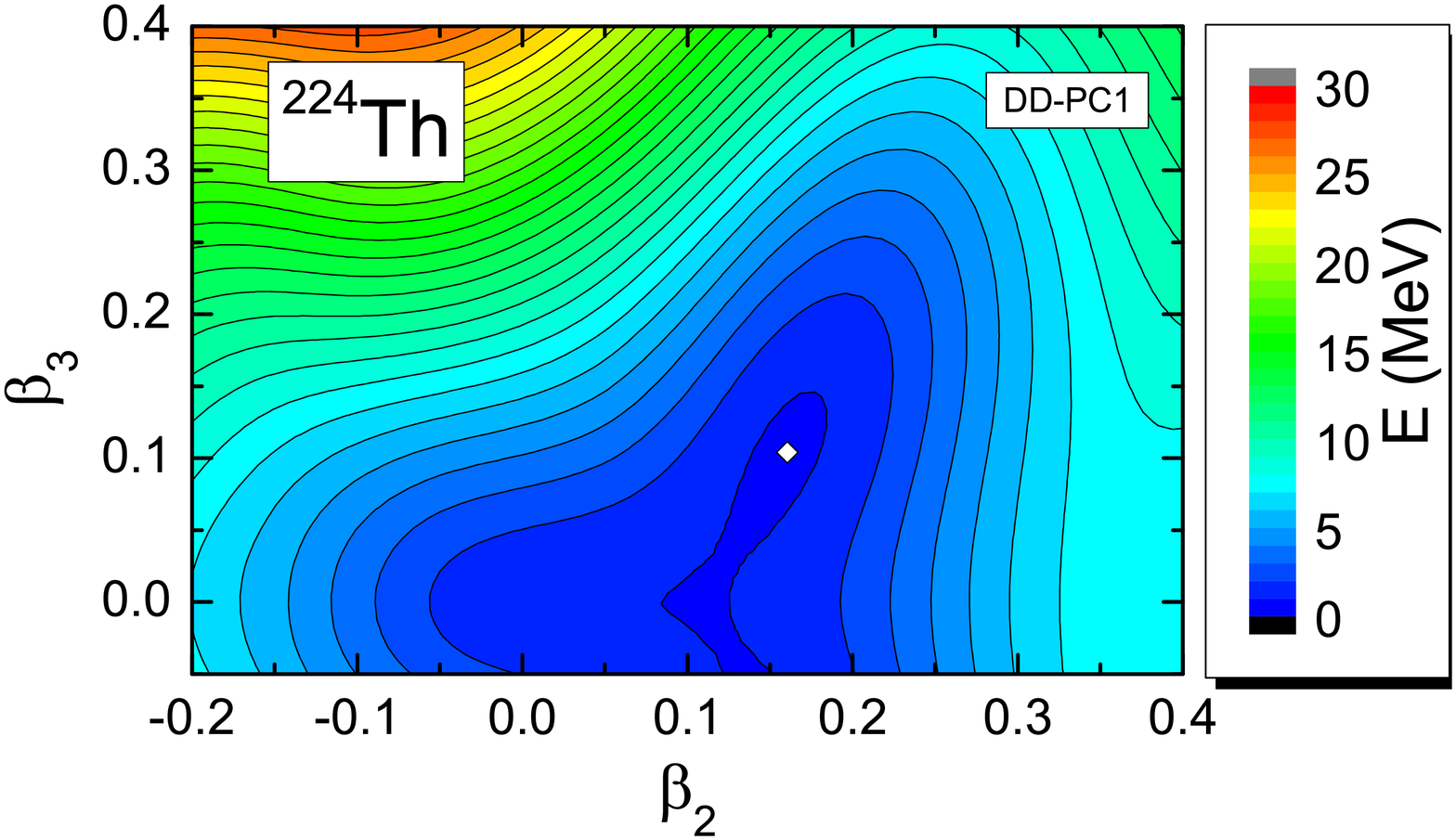} &
\includegraphics[width=5.7cm]{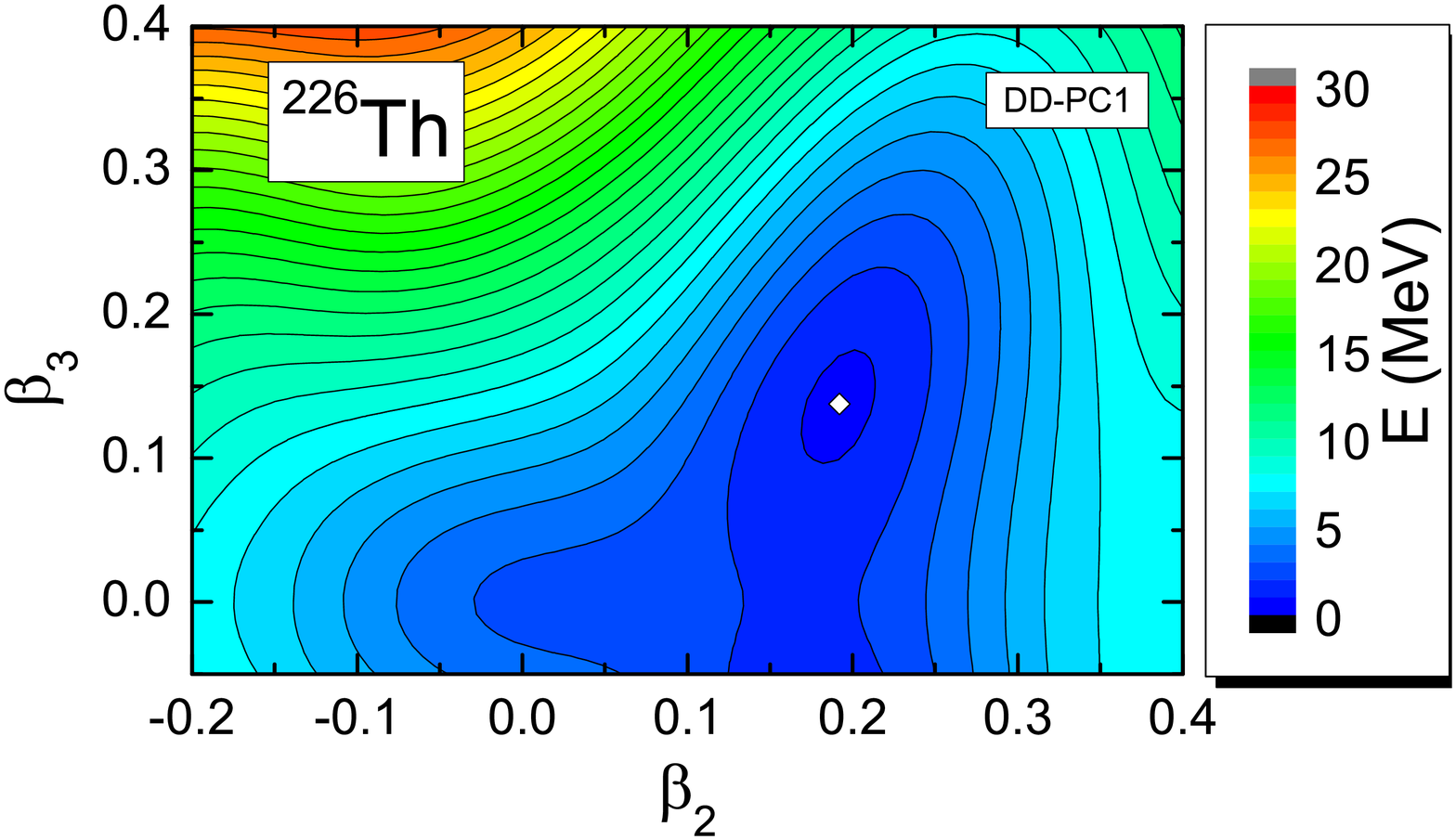} \\
\includegraphics[width=5.7cm]{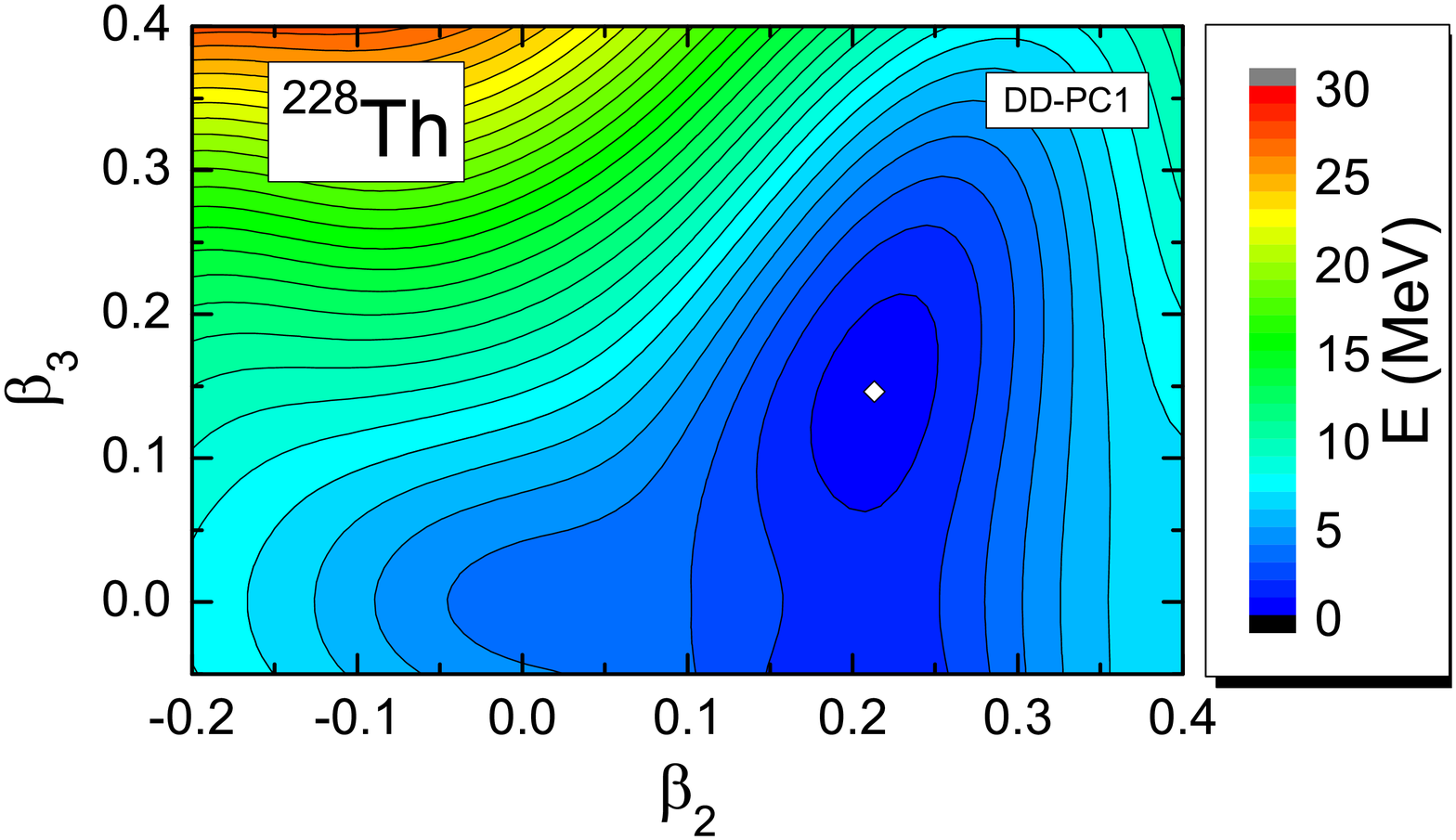} &
\includegraphics[width=5.7cm]{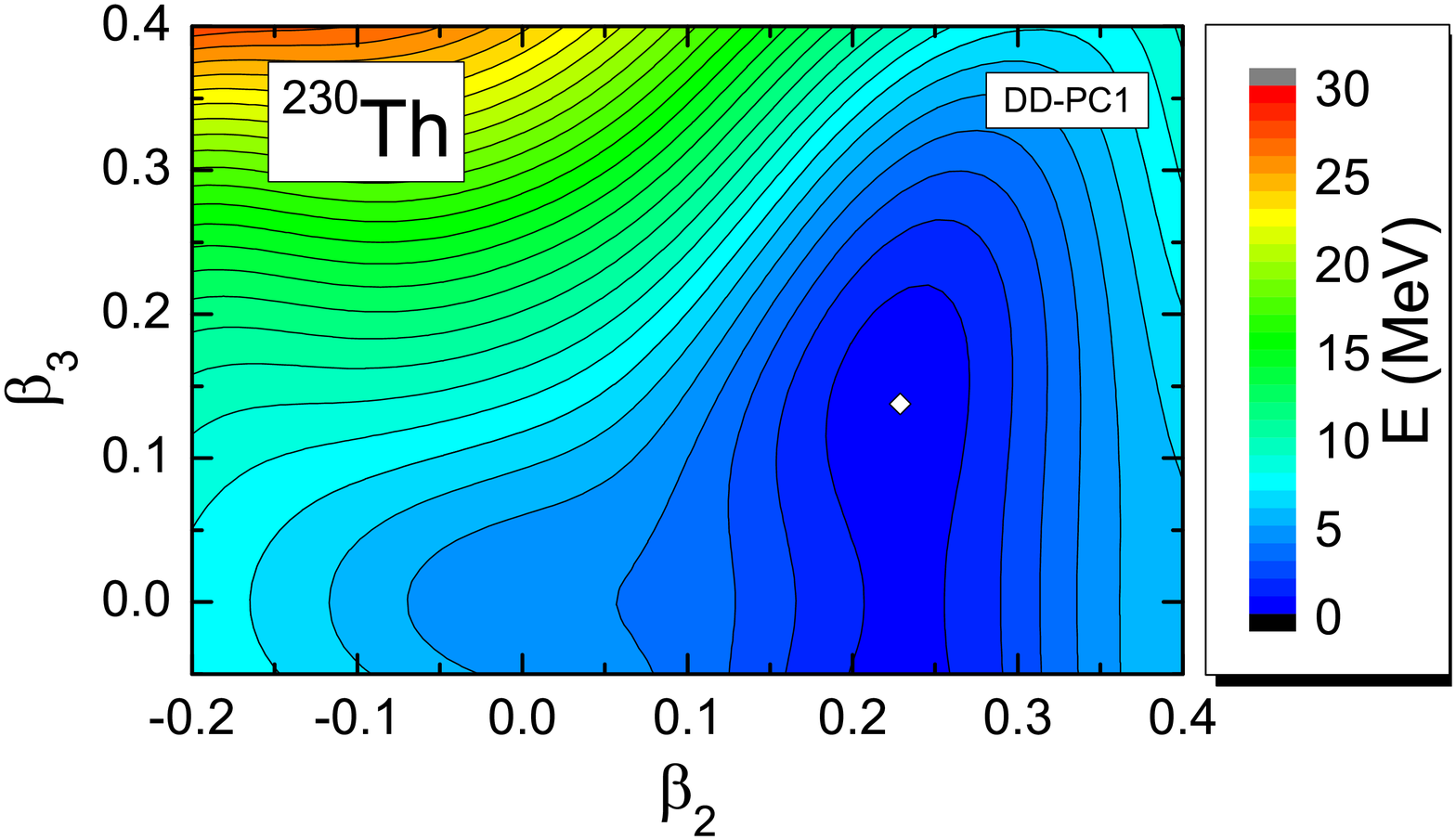} &
\includegraphics[width=5.7cm]{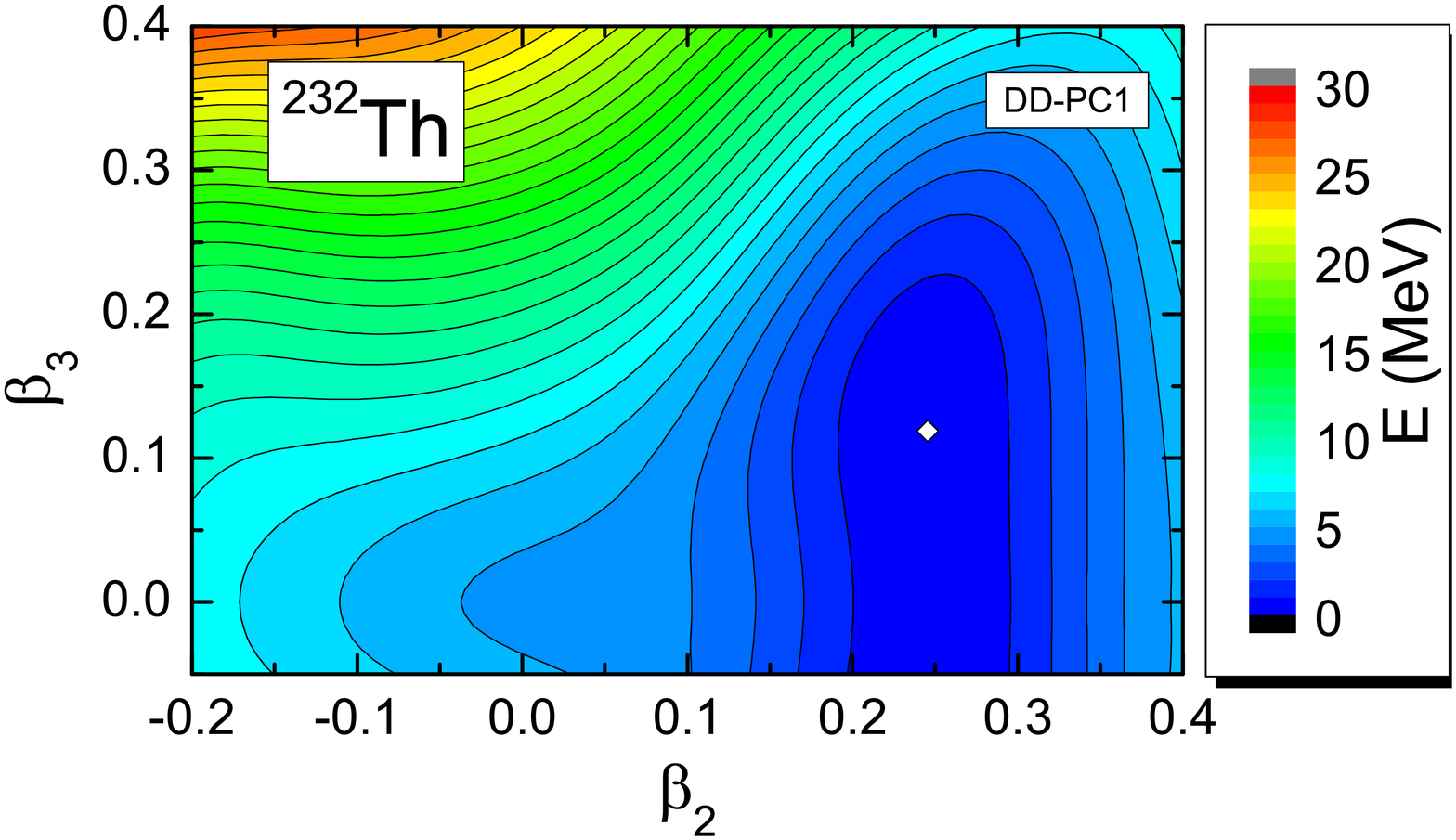} \\
\end{tabular}
\caption{(Color online) Microscopic DD-PC1 self-consistent RHB 
axially symmetric energy surfaces of the nuclei $^{222-232}$Th in the $(\beta_{2},\beta_{3})$
 plane, symmetric with respect to $\beta_{3}=0$ axis. The contours join points on the surface with the 
 same energy, and the separation between neighboring contours is 1
 MeV. In each panel, the minimum is indicated by a open diamond. }
\label{fig:pes}
\end{center}
\end{figure*}

At present the most comprehensive approach to nuclear structure is based on the framework 
of energy density functionals (EDFs). Nuclear EDFs enable a complete and accurate description 
of ground-state properties and collective excitations over the whole nuclide chart \cite{ben03rev}. 
Both non-relativistic  and relativistic
EDFs have successfully been applied to the description of the evolution 
of single-nucleon shell-structures and the related shape-transition and shape-coexistence 
phenomena. The calculations reported in this work are based on the relativistic EDF
DD-PC1  \cite{DDPC1}. This functional has been  
employed in a number of mean-field and beyond-mean-field studies  
of structure phenomena in various mass regions \cite{Nik11rev}, 
from the evolution of shapes in $N=28$ isotones \cite{DDPC1-N28},  
to rapid shape transitions in superheavy nuclei \cite{DDPC1-SHE}. 
Rather than using a specifically designed potential model that by construction describes the critical point 
of octupole phase transition \cite{bizzeti04,bonatsos05,lenis06,bizzeti08,bizzeti10}, 
here we employ a global density functional that was not specifically adjusted nor ever before 
applied to studies of octupole shapes and negative-parity states. 

The analysis starts by performing
constrained self-consistent relativistic mean-field calculations for 
axially symmetric shapes in the ($\beta_{2}$,$\beta_{3}$) plane, with 
constraints on the mass quadrupole $Q_{2 0}$, and octupole $Q_{3 0}$ moments. 
The dimensionless shape variables $\beta_{\lambda}$ ($\lambda=2,3$) are defined in terms of 
the multipole moments $Q_{\lambda 0}$: $\beta_{\lambda}\equiv ({4\pi}/{3AR^{\lambda}})
Q_{\lambda 0}$, with $R=1.2A^{1/3}$ fm. 
The relativistic Hartree-Bogoliubov (RHB) model \cite{Vre05} is used to calculate 
constrained energy surfaces (cf. \cite{lu13} for details), the 
functional in the particle-hole channel is DD-PC1 and pairing correlations are taken into account 
by employing an interaction that is separable in momentum space, and is completely determined 
by two parameters adjusted to reproduce the empirical bell-shaped pairing gap in symmetric 
nuclear matter \cite{tian09,Nik11rev}. 

Figure \ref{fig:pes} displays the contour plots of deformation energy surfaces in the
($\beta_{2}$,$\beta_{3}$) plane for the isotopes $^{222-232}$Th. The plots are
symmetric with respect to the $\beta_{3}=0$ axis. Already at 
the mean-field level the RHB model predicts a very interesting structural evolution. 
A soft energy surface is calculated for $^{222}$Th, with the energy minimum close to 
$(\beta_{2},\beta_{3})\approx (0,0)$. 
The quadrupole deformation becomes more pronounced in $^{224}$Th, and one also notices 
the development of octupole deformation. The energy minimum is found in the 
$\beta_{3}\neq 0$ region, located at $(\beta_{2},\beta_{3})\approx (0.15,0.1)$. 
From $^{224}$Th to $^{226,228}$Th a rather strongly marked octupole minimum 
is predicted. The deepest octupole minimum is calculated in $^{226}$Th 
whereas, starting from $^{228}$Th, the minimum becomes softer in $\beta_{3}$ direction. 
Soft octupole surfaces are obtained for $^{230,232}$Th, the latter being completely flat in $\beta_{3}$. 

A quantitative study of shape transitions must go beyond a simple mean-field 
calculation of potential energy surfaces and, particularly in the case of a possible 
QPT, it must include the computation of observables that can be related to quantum order 
parameters. In this work we employ the interacting boson model (IBM) \cite{IBM} to 
calculate spectroscopic properties
associated to quadrupole and octupole deformations. 
The building blocks of the IBM include the monopole $s$ and the quadrupole $d$ bosons, 
corresponding to collective $J^{\pi}=0^{+}$ and $2^{+}$ pairs of
valence nucleons, respectively \cite{OAI}. 
To describe reflection-asymmetric deformations and the corresponding
negative-parity states, in addition to the positive-parity bosons, the model space must 
include the octupole ($J^{\pi}=3^{-}$) boson $f$ \cite{engel87}.  Here 
we employ the following $sdf$-IBM Hamiltonian similar to the
one used in \cite{barfield88}: 
\begin{eqnarray}
\label{eq:ham}
\hat H=\epsilon_{d}\hat n_{d}+\epsilon_{f}\hat n_{f}+\kappa_{2}\hat
 Q\cdot\hat Q+\alpha\hat L_{d}\cdot\hat L_{d}+\kappa_{3}:\hat V_{3}^{\dagger}\cdot\hat V_{3}:
\end{eqnarray}
where $\hat n_{d}=d^{\dagger}\cdot\tilde d$  and $\hat
n_{f}=f^{\dagger}\cdot\tilde f$ denote the $d$ and $f$ boson
number operators, respectively. The third term is the quadrupole-quadrupole interaction 
with the quadrupole operator $\hat Q=s^{\dagger}\tilde
d+d^{\dagger}s+\chi_{d}[d^{\dagger}\times\tilde
d]^{(2)}+\chi_{f}[f^{\dagger}\times\tilde f]^{(2)}$. The angular momentum operator
in the $sd$ space reads $\hat L_{d}=\sqrt{10}[d^{\dagger}\times\tilde d]^{(1)}$, and 
the last term in Eq.~(\ref{eq:ham}) denotes a specific octupole-octupole
interaction expressed in normal-ordered form with $\hat
V_{3}^{\dagger}=s^{\dagger}\tilde f+\chi_{3}[d^{\dagger}\times\tilde
f]^{(3)}$. 

\begin{figure}[ctb!]
\begin{center}
\includegraphics[width=8.2cm]{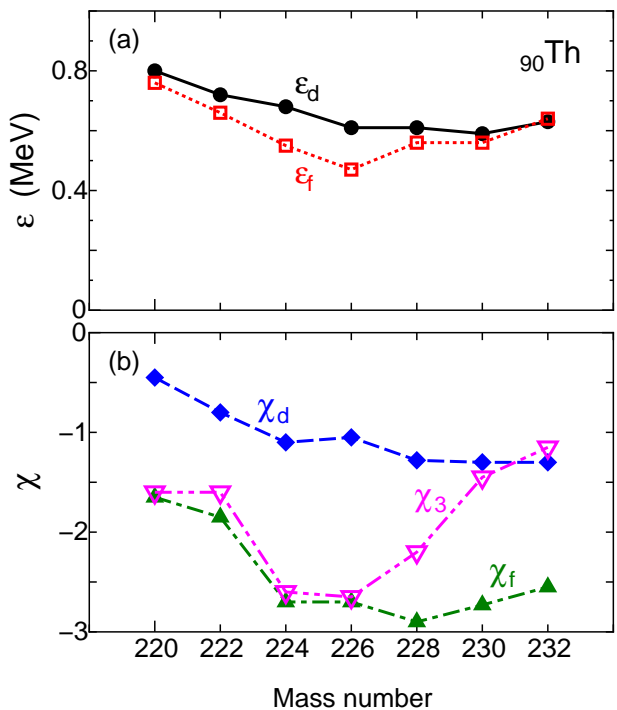}
\caption{(Color online) Variation of the derived parameters
 $\epsilon_{d}$, $\epsilon_{f}$, $\chi_{d}$, $\chi_{f}$ and $\chi_{3}$ as
 functions of the mass number. }
\label{fig:para}
\end{center}
\end{figure}

For each nucleus the Hamiltonian parameters: $\epsilon_{d}$,
$\epsilon_{f}$, $\alpha$, $\kappa_{2}$,
$\kappa_{3}$, $\chi_{d}$, $\chi_{f}$ and $\chi_{3}$, are determined by employing the procedure 
of Ref.~\cite{Nom08}: the microscopic self-consistent
mean-field energy surface is mapped onto the equivalent IBM energy
surface, that is, on the expectation value of the IBM Hamiltonian 
$\langle\phi|\hat H|\phi\rangle$ in the boson 
condensate state $|\phi\rangle$ \cite{GK} (see Refs.~\cite{Nom08,Nom10} for details).  
$|\phi\rangle=\frac{1}{\sqrt{N_{B}!}}(\lambda^{\dagger})^{N_{B}}|-\rangle$, with
$\lambda^{\dagger}=s^{\dagger}+\beta_{2}d^{\dagger}_{0}+\beta_{3}f^{\dagger}_{0}$.  
$N_{B}$ and $|-\rangle$ denote the number of bosons, that is, half the number of
valence nucleons \cite{OAI}, and the boson vacuum (a core with
doubly-closed shells), respectively. In the present case 
the doubly-magic nucleus $^{208}$Pb plays the role of the boson vacuum. 
Thus, $N_{B}$ varies between 6 and 12 for the $^{220-232}$Th nuclei. 
By equating the expectation value of the $sdf$ IBM Hamiltonian as function of 
$\beta_{2}$ and $\beta_{3}$ to the microscopic energy surface in the
neighborhood of the minimum, the Hamiltonian parameters can be
determined without invoking any further adjustment to data. 
Once the parameters are specified, the Hamiltonian of
Eq.~(\ref{eq:ham}) is numerically diagonalized by using the code
OCTUPOLE \cite{OCTUPOLE} to generate energy spectra and transition
rates. 

In Fig.~\ref{fig:para} we plot the microscopically determined values of the 
IBM Hamiltonian parameters $\epsilon_{d}$,
$\epsilon_{f}$, $\chi_{d}$, $\chi_{f}$ and $\chi_{3}$ for 
Th isotopes, as functions of the mass number. The decrease of the 
$d$-boson energy $\epsilon_{d}$ in Fig.~\ref{fig:para}(a) reflects the 
enhancement of quadrupole collectivity in heavier Th isotopes (cf. Fig.~\ref{fig:pes}). 
$\epsilon_{d}$ is nearly constant from $^{226}$Th up to $^{230}$Th. 
The $f$-boson energy $\epsilon_{f}$ also decreases from $^{220}$Th to $^{226}$Th, 
and from that isotope its value increases with mass.
An interesting feature to be noted in Fig.~\ref{fig:para}(a) is that  $\epsilon_{f}$
is of the same order of magnitude as $\epsilon_{d}$, and this implies that the 
octupole deformation can be as pronounced as the quadrupole one. 
In fact, the octupole minimum on the RHB energy surfaces is rather deep
for $A\geq 226$ (Fig.~\ref{fig:pes}). We note that, in contrast,  
most phenomenological $sdf$-IBM studies have assumed a
rather weak coupling between positive and negative-parity bosons,
$\epsilon_{d}\ll\epsilon_{f}$ \cite{engel87}.

Figure~\ref{fig:para}(b) shows that the quadrupole parameter $\chi_{d}$ increases in magnitude with
$A$, and its value is close to the SU(3) limit of the $sd$ IBM
$\chi_{d}=-\sqrt{7}/2\approx 1.3$ \cite{IBM} for $A\geq 228$. 
In the same panel, both $\chi_{f}$ and $\chi_{3}$ exhibit a
rapid change from $^{222}$Th to $^{224}$Th, corresponding to
the onset of quadrupole and octupole deformations. Starting 
from $^{226}$Th, $\chi_{f}$ is rather constant, whereas the value of 
$\chi_{3}$ decreases in magnitude reflecting the softness of the  octupole
minimum (Fig.~\ref{fig:pes}). 

Nearly constant values are adopted for the remaining strength parameters:
$\kappa_{2}\approx -0.06$ MeV and $\kappa_{3}\approx -0.015$ MeV. 
The coupling constant of the $\hat L_{d}\cdot\hat L_{d}$ term $\alpha\approx -0.02$ MeV 
is determined separately so that the cranking
moment of inertia in the IBM intrinsic state, corresponding to the
minimum on the $\beta_{2}$ axis, becomes identical to the 
one computed in the mean-field model \cite{Nom11rot}. 

\begin{figure}[ctb!]
\begin{center}
\includegraphics[width=8.2cm]{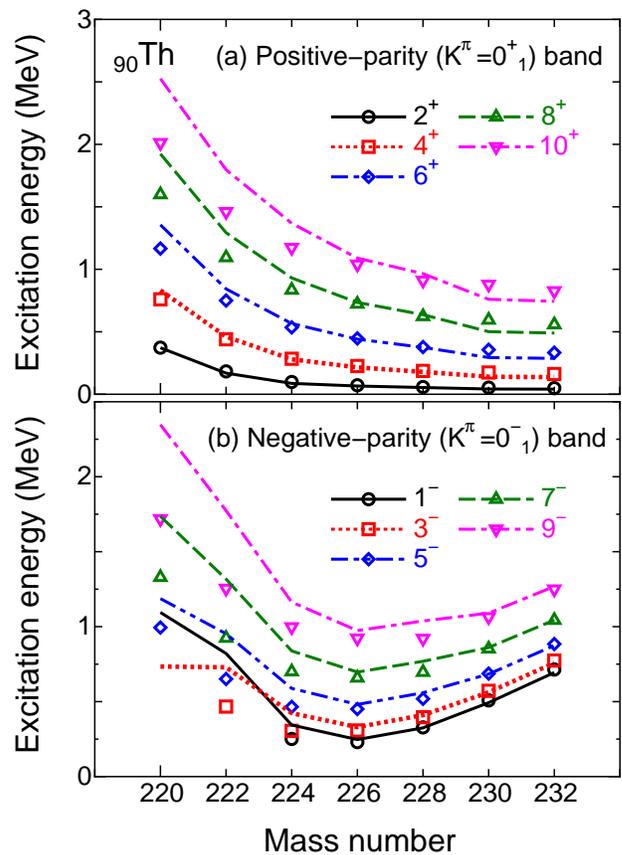}
\caption{(Color online) Isotopic dependence 
of the excitation energies of levels of the positive-parity
ground-state band ($K^{\pi}=0^{+}_{1}$) (a), and the lowest
negative-parity band ($K^{\pi}=0^{-}_{1}$) (b), for $^{220-232}$Th.
In each panel lines and symbols denote the theoretical and the experimental
 \cite{data} values, respectively. }
\label{fig:spectra}
\end{center}
\end{figure}

A signature of stable octupole deformation is a low-lying negative-parity band 
$J^\pi =$ $1^-, 3^-, 5^-, \ldots$ located close in energy to the positive-parity 
ground-state band $J^\pi =$ $0^+, 2^+, 4^+, \ldots$, thus forming an
alternating-parity band. Such alternating bands are 
typically observed for states with spin $J\geq 5$ \cite{butler96}. 
In the case of octupole vibrations the negative-parity band is found at 
higher energy, and the two sequences of positive- and negative-parity 
states form separate collective bands.
Therefore, a systematic increase with nucleon 
number of the energy of the negative-parity band 
relative to the positive-parity ground-state band indicates 
a transition from stable octupole deformation to octupole vibrations
\cite{butler96}. 

In Fig.~\ref{fig:spectra} we display the systematics of 
calculated excitation energies of the ground-state band 
($K^{\pi}=0^{+}_{1}$) 
and the lowest negative-parity band ($K^{\pi}=0^{-}_{1}$) in $^{220-232}$Th, in 
comparison to available data \cite{data}. 
For all isotopes these two bands are formed by zero $f$-boson and one $f$-boson states,
respectively. Even without any adjustment  to the data, that is, 
by simply using parameters determined by the 
microscopic calculation of potential energy surfaces, the IBM quantitatively 
reproduces the isotopic dependence of excitation energies of levels 
belonging to the lowest bands of positive and negative parity. 

Positive-parity levels, shown in Fig.~\ref{fig:spectra}(a), systematically decrease in energy 
with mass number, reflecting the increase of quadrupole collectivity
(cf. Fig.~\ref{fig:pes}). 
$^{220,222}$Th exhibit a quadrupole vibrational structure, whereas 
pronounced ground-state rotational bands with
$E(4^{+}_{1})/E(2^{+}_{1})\approx 10/3$ are found in $^{226-232}$Th. 

In Fig.~\ref{fig:spectra}(b) the calculated excitation energies of the negative-parity band form a parabolic 
structure centered between $^{224}$Th and $^{226}$Th. 
The approximate parabola of $1^{-}_{1}$ states displays a minimum 
at $^{226}$Th, in which the octupole deformed minimum 
is most pronounced (cf. Fig. \ref{fig:pes}), in agreement with 
the mass dependence of the experimental
$1^{-}_{1}$ level: $E(1^{-}_{1})=$ 251, 230 and 328 keV in
$^{224,226,228}$Th nuclei, respectively \cite{data}. 
Starting from $^{226}$Th, the energies of negative-parity 
states systematically increase and the band becomes more 
compressed. A rotational-like collective band based on the octupole vibrational
$1^{-}_{1}$ state, that is, on the state that corresponds to non-static octupole
deformation, develops. This result correlates with the systematics of 
microscopic energy surfaces that become softer in $\beta_{3}$ starting from
$^{226}$Th (Fig.~\ref{fig:pes}), and with the 
increase (decrease) of the parameter $\epsilon_{f}$ ($|\chi_{3}|$) shown in
Fig.~\ref{fig:para}(a) ((b)). 
The calculated negative-parity states for the lightest nuclei $^{220,222}$Th
are somewhat higher in energy when compared to the data
(Fig.~\ref{fig:spectra}(b)). 
The reason is that the valence space may not be large enough for these nuclei. 

The vertex of the parabola of the calculated negative-parity states 
(Fig.~\ref{fig:spectra}(b)) can be associated 
with a QPT between stable octupole deformations and octupole 
vibrations characteristic for $\beta_3$-soft potentials, with the 
excitation energy of the negative-parity band (e.g., the $1^{-}_{1}$
bandhead energy) representing the order parameter  
for this shape transition. Phenomenological studies (e.g.,
\cite{bizzeti04,bonatsos05,lenis06,bizzeti08})  
were only able to reproduce this QPT by employing Hamiltonians in which the degree of 
quadrupole-octupole correlations is controlled by a model parameter adjusted 
to  the empirical position of the critical point. 
In contrast, the framework used in this work provides a fully microscopic prediction of the QPT. 
The shape phase transition occurs as a function of the physical control 
parameter -- the nucleon (neutron) number, and the isotope $^{226}$Th 
is found to be closest to the critical point. As we have already 
emphasized, in the case of atomic nuclei the control parameter of 
shape QPTs is discrete and, therefore, it is not always possible 
to associate a specific isotope to the critical point.

\begin{figure}[ctb!]
\begin{center}
\includegraphics[width=8.5cm]{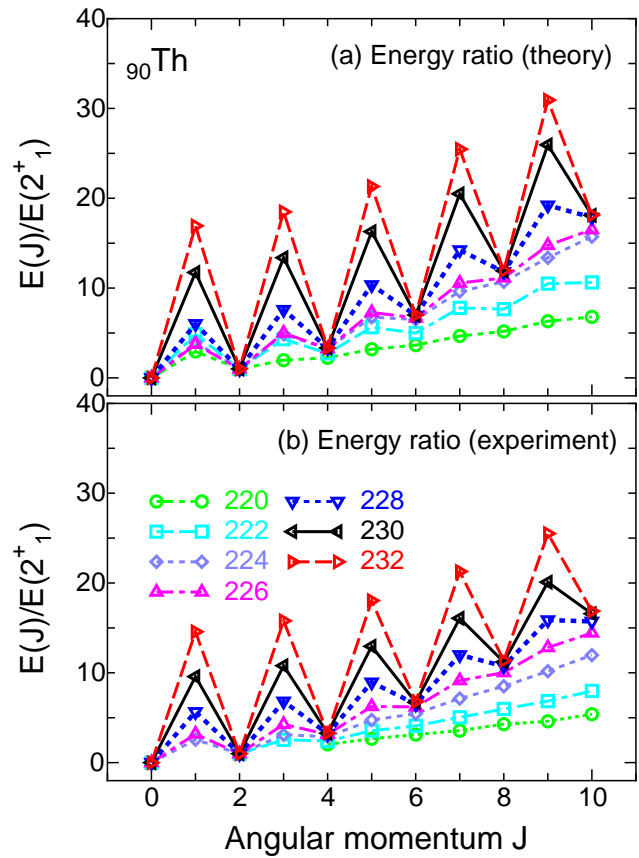}
\caption{(Color online)  Theoretical (a) and experimental \cite{data} (b) 
energy ratios $E(J)/E(2^+_1)$ of the yrast states of $^{220-232}$Th, 
including both positive ($J$ even)
and negative ($J$ odd) parity, as 
functions of the angular momentum $J$.  
}
\label{fig:ratios}
\end{center}
\end{figure}

Another indication of 
the phase transition between octupole  
deformation and octupole vibrations for $\beta_3$-soft potentials 
is provided by the odd-even staggering in the energy ratio
$E(J)/E(2^+_1)$. 
Figure~\ref{fig:ratios} displays the ratios $E(J)/E(2^+_1)$ for both
positive- and negative-parity yrast states of $^{220-232}$Th as 
functions of the angular momentum $J$. 
Below $^{226}$Th the odd-even staggering is negligible, indicating that positive and 
negative parity states are lying close to each other in energy.  
The staggering only 
becomes more pronounced starting from $^{228}$Th, and this 
means that negative-parity states form a separate rotational band 
built on the octupole vibration. 
In particular, the energy ratio $E(J)/E(2^+_1)$ for negative-parity
(odd-$J$) states could be considered as  
an order parameter for the octupole shape transition. 
We note that the predicted staggering 
of yrast states is in very good agreement with data \cite{data}.

\begin{figure}[ctb!]
\begin{center}
\includegraphics[width=8.6cm]{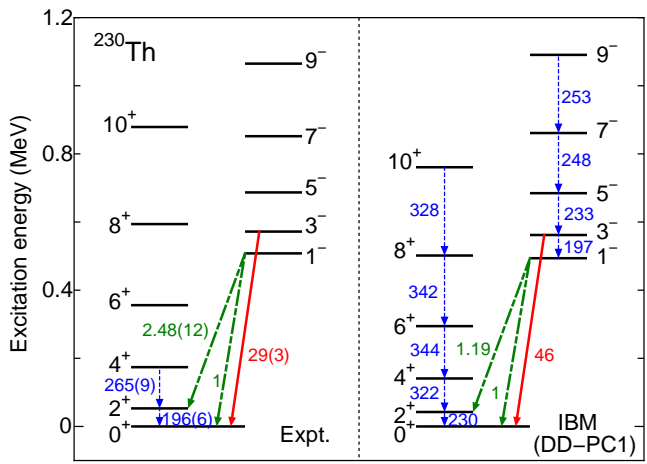}
\caption{(Color online)  Experimental  \cite{data,ackermann93} and calculated 
yrast states of positive and negative parity in $^{230}$Th. 
The in-band $B$(E2) values (dotted) and 
the $B({\textnormal{E3}};3^{-}_{1}\rightarrow 0^{+}_{1})$ (solid)
(both in Weisskopf units), 
and the branching ratio $B({\textnormal{E1}};1^{-}_{1}\rightarrow 2^{+}_{1})/B
({\textnormal{E1}};1^{-}_{1}\rightarrow 0^{+}_{1})$ (dashed-dotted) are also shown.}
\label{fig:230Th}
\end{center}
\end{figure}

To illustrate in more detail the level of quantitative agreement between our 
microscopic model calculation and data, 
in Fig.~\ref{fig:230Th} we display the energy spectrum of positive and 
negative parity yrast states of the 
octupole-soft nucleus $^{230}$Th, including the in-band $B$(E2) values and 
the $B({\textnormal{E3}};3^{-}_{1}\rightarrow 0^{+}_{1})$ (both in Weisskopf units), 
and the branching ratio $B({\textnormal{E1}};1^{-}_{1}\rightarrow 2^{+}_{1})/B
({\textnormal{E1}};1^{-}_{1}\rightarrow 0^{+}_{1})$. 
The E1, E2 and E3 operators read
$\hat T^{{\textnormal{(E1)}}}=e_{1}(d^{\dagger}\times\tilde f+f^{\dagger}\times\tilde
d)^{(1)}$, $\hat T^{{\textnormal{(E2)}}}=e_{2}\hat Q$ and
$\hat T^{{\textnormal{(E3)}}}=e_{3}(\hat V_{3}^{\dagger}+\hat V_{3})$, respectively, 
with the effective charges 
$e_{2}=0.19$ $e$b and $e_{3}=0.19$ $e$b$^{3/2}$ 
taken from previous empirical studies in Refs. \cite{taka88} and
\cite{cottle98}, respectively. One notices a very good  
agreement with experiment \cite{data,ackermann93}, not only for excitation energies 
but also for transition probabilities. 

Finally, we note that the connection between the 
evolution of collective excitations and the occurrence 
of a nuclear shape QPT has, in many studies, been investigated using 
symmetry-dictated approaches, including the IBM. 
In these studies the concept of a QPT is closely related to the group
structure of a schematic IBM Hamiltonian, that is, a Hamiltonian 
of Ising type \cite{cejnar10rev}. Since the number of shape degrees of 
freedom becomes rather large when octupole
deformation is taken into account, a major challenge is to develop 
a symmetry-dictated approach that captures the physics of both quadrupole and
octupole collective degrees of freedom, thereby providing a phase diagram in the 
parameter space associated to a certain symmetry structure of the Hamiltonian.  
In the present work, on the other hand, the Hamiltonian in Eq.~(\ref{eq:ham}) 
takes the simplest possible form and the parameters are  
determined solely from the basic topology of the microscopic energy surface
in the vicinity of its minimum and, therefore, no symmetry structure is imposed on
the Hamiltonian. The development of a symmetry-dictated description of 
shape phase transitions considered in this work presents an interesting 
problem for future studies.

In summary, we have performed a microscopic analysis of a transition between stable octupole
deformation and octupole vibrations in Th isotopes. 
A global relativistic EDF, not specifically adjusted to  
octupole degrees of freedom nor ever before applied to 
reflection-asymmetric nuclei, has been used to calculate 
axially-symmetric constrained energy surfaces in the 
($\beta_{2}$,$\beta_{3}$) plane. 
The $sdf$ IBM Hamiltonian has been constructed by
mapping the microscopic energy surface onto the equivalent one in the
boson system, providing the
low-energy excitation spectra and transition rates,  
that is, observables that can be related to quantum order parameters. 
%
The microscopic model predicts a transition from spherical shapes 
near $^{220}$Th to stable octupole and quadrupole deformations around
$^{226}$Th, and the development of octupole vibrations  
characteristic for $\beta_3$-soft potentials in heavier Th nuclei. 
The EDF constrained microscopic energy surfaces (Fig. \ref{fig:pes}) and 
the systematics of low-energy excitation spectra (Figs.~\ref{fig:spectra} 
and \ref{fig:ratios}) point to the occurrence of a shape phase transition 
near $^{226}$Th. With increasing neutron number the octupole deformation 
appears to be $\beta_3$-unstable (soft) and remains such up to $^{232}$Th.
This result is in excellent agreement with available 
data and previous phenomenological studies of phase 
transitions in octupole collective degrees of freedom. Based on 
a comparison with previous work, it appears that Th presents the 
best case for octupole deformations in atomic nuclei.

The authors would like to thank R. V. Jolos and J. Zhao for useful discussions. 
K. N. acknowledges support by the JSPS Postdoctoral
Fellowships for Research Abroad. 
Calculations were partly performed 
on the ScGrid of the 
Supercomputing Center, Computer Network Information Center of Chinese Academy of Sciences.

\bibliography{refs}

\end{document}